\documentclass[aps, prb, reprint]{revtex4-1}
\usepackage{bm}
\usepackage{amsmath}
\usepackage{graphicx}

\def\Vec#1{\mbox{\boldmath $#1$}}

\begin{document}

\title{Critical exponent of metal-insulator transition
  in doped semiconductors: \\
  the relevance of the Coulomb interaction}

\date{\today}

\author{Yosuke Harashima}
\altaffiliation{Present address:
  Elements Strategy Initiative Center for Magnetic Materials,
  National Institute for Materials Science,
  1-2-1 Sengen,
  Tsukuba,
  Ibaraki 305-0047,
  Japan}
\author{Keith Slevin}
\affiliation{Department of Physics,
  Graduate School of Science,
  Osaka University,
  1-1 Machikaneyama,
  Toyonaka,
  Osaka 560-0043,
  Japan}

\begin{abstract}
  We report a simulation of the metal-insulator transition in a model of a doped semiconductor that treats disorder and interactions on an equal footing.
  The model is analysed using density functional theory. From a multi-fractal analysis of the Kohn-Sham
  eigenfunctions, we find $\nu \approx 1.3$ for the critical exponent of the correlation length.
  This differs from that of Anderson's model of localisation and suggests that the Coulomb interaction changes the universality class of the transition.
\end{abstract}

\maketitle

\section{Introduction}

In heavily doped semiconductors a zero temperature metal-insulator transition (MIT) is observed
as a function of doping concentration $N_{\text{d}}$.
For samples with doping concentrations below a critical concentration $N_{\text{c}}$,
the conductivity extrapolated to zero temperature $\sigma(T=0)$ is found to be zero,
while for samples with concentrations exceeding this critical concentration,
$\sigma(T=0)$ is finite.
Well studied examples are the transition in Si:P
\cite{rosenbaum80,stupp93,lohneysen11}
and Ge:Ga \cite{itoh04} and there are many others
(see Ref. \onlinecite{itoh04} and references therein).
The observed critical concentrations obey, approximately at least, the relation\cite{edwards78}
\begin{equation} \label{eq:mott-criterion}
  N_{\text{c}}^{\frac{1}{3}} a_{B}^{*} \approx 0.26\;.
\end{equation}
Here, $a_{B}^{*} \equiv (\varepsilon_{r} / m^*) a_{B}$
is the effective Bohr radius of a hydrogenic impurity state
for a carrier with effective mass $m^*$ in a medium with
relative dielectric constant $\varepsilon_{r}$.

The transition is continuous and considerable effort has been expended to study
the critical phenomena
observed at the transition and the associated critical exponents, the values of which
are expected to be universal.
As the transition is approached from higher concentration $N_{\text{d}}>N_{\text{c}}$ it
is found that the zero temperature conductivity obeys a power law with a
critical exponent $\mu$
\begin{equation}\label{eq:zeroT}
  \sigma(N_{\text{d}}, T=0) \propto \left( N_{\text{d}} - N_{\text{c}} \right)^{\mu}\;.
\end{equation}
Analyses have also been performed which avoid the extrapolation to zero temperature
and fit data at finite temperatures directly to the dynamic scaling law\cite{wegner76}
\begin{equation}\label{eq:finiteT}
  \sigma(N_{\text{d}},T) = T^x f\left( \left(N_{\text{d}}/N_{\text{c}} - 1\right) T^{-y} \right)
\end{equation}
Here, $x=1/z$ and $y=1/(z\nu)$ with $\nu$ the critical exponent describing the divergence of
the correlation length at the transition and $z$ the dynamic exponent.
Demanding consistency between these scaling laws gives
$\mu=\nu$, which is known as Wegner's scaling relation.

For Si:P there has been debate over the value of the critical
exponent; $\mu\approx 0.5$ or $\mu\approx 1.3$\cite{rosenbaum94, stupp94}.
Itoh et al addressed this controversy by a careful study of the
MIT in neutron transmutation doped Ge:Ga \cite{itoh04}.
They demonstrated that the width of the critical region depends on the compensation.
For intentionally compensated samples they found
both Eq. (\ref{eq:zeroT}) and dynamic scaling Eq. (\ref{eq:finiteT}) are observed
with $\mu\approx 1$ and $z\approx 3$.
In nominally uncompensated samples, Eq. (\ref{eq:zeroT}) with $\mu \approx 0.5$ is observed
over a wide range of concentrations on the metallic side but dynamic scaling is violated.
However, when attention is focused on a much narrower region around the critical point,
Eq. (\ref{eq:zeroT}) with $\mu \approx 1$ is a better fit and dynamic scaling is recovered with
$\mu\approx 1$ and $z\approx 3$.
Dynamic scaling with similar values of the exponents has also been reported in Si:P\cite{lohneysen11}.

There is still no explanation of the
critical behaviour described above.
Since the impurities used to dope the semiconductor
are randomly distributed in space, an Anderson transition is a possibility.
However, the value of the critical exponent is not consistent with this.
While $z=3$ is expected for Anderson's model of localisation since
the only relevant energy scale at
the transition is the level spacing, numerical finite size scaling of this model
has established that
$\nu = 1.57 \pm .02$\cite{slevin99}.
This estimate has been confirmed in subsequent work
\cite{markos06, garcia07, rodriguez10, rodriguez11}.
Clear demonstrations of the universality of this value include its confirmation
in a non-interacting tight-binding model with random hopping of a doped semiconductor\cite{krich11}, and its
experimental measurement in a quasi-periodic atomic kicked rotor\cite{lopez12}.
This suggests that an explanation of the critical behaviour must incorporate
both electron-electron interactions and disorder.
While there have been serious efforts to develop such a theory
(see [\onlinecite{finkelstein10}] and [\onlinecite{vlad12}]
for an up to date discussion),
it remains one of the most challenging problems in condensed matter physics.

\begin{figure}[h]
  \begin{center}
    \includegraphics[width = \hsize]{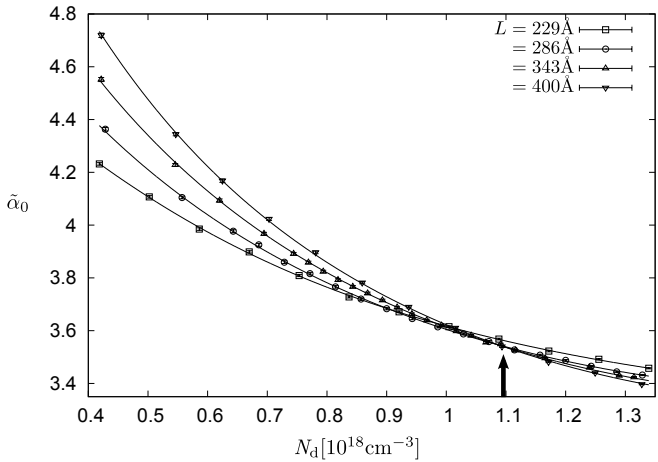}
  \end{center}
  \caption{The generalized multi-fractal exponent $\tilde{\alpha}_{0}$ as a function of
    donor concentration for several system sizes.
    The solid lines are a finite size scaling fit to the data.
    The critical concentration is indicated by an arrow.}
  \label{fig:alpha0}
\end{figure}

The object of this paper is to take a step closer to an understanding of the MIT in
doped semiconductors by determining how the Coulomb interaction affects the critical
behaviour of the Anderson transition.
We do this by simulating a model of a doped
semiconductor that treats on an equal footing both the disorder due to the random spatial distribution of the
dopants and the Coulomb interaction between the carriers.
Applying a multi-fractal finite size scaling method\cite{rodriguez10,
rodriguez11}, we find that the model exhibits a localization-delocalization
transition at approximately the right carrier concentration (see Figure \ref{fig:alpha0}).
Moreover, we find that the critical exponent $\nu\approx 1.3$, which is
different from that for the standard Anderson transition.

\section{Model and Simulation Method}

The main difficulty in a numerical study of this problem is to
find a model that is numerically tractable but still captures the physics of
both disorder and interactions.
The wide applicability of the Mott criterion suggests that a good starting point
is to treat the semiconductor
as an effective medium with appropriate dielectric constant
and effective mass.
The donors are modelled as unit positive charges randomly distributed in
this medium.
The Hamiltonian (in Hartree atomic units) is\cite{harashima12}
\begin{eqnarray}
  \mathcal{H}
  & = &
  - \frac{1}{2 m^{*}} \sum_{i=1}^{N} \nabla_{i}^{2}
  - \frac{1}{\varepsilon_{r}} \sum_{i, I=1} ^{N}
  \frac{1}{\left| \Vec{r}_{i} - \Vec{R}_{I} \right|}
  \nonumber \\
  &&
  + \frac{1}{2 \varepsilon_{r}}
  \sum_{i \neq j}^{N} \frac{1}{\left| \Vec{r}_{i} - \Vec{r}_{j} \right|}
  + \frac{1}{2 \varepsilon_{r}}
  \sum_{I \neq J}^{N} \frac{1}{\left| \Vec{R}_{I} - \Vec{R}_{J} \right|}
  \label{eq:hamiltonian}
\end{eqnarray}
Here, $\{ \Vec{R}_{I} \}$ are the random positions of the donors and
$N$ is the number of donors.
Each impurity donates a single electron so the number of electrons is also $N$
and the total charge is zero.

The properties of the model scale exactly with
the dielectric constant and effective mass.
By rescaling lengths according to
\begin{equation}
  \tilde{\Vec{r}} = \left( m^* / \varepsilon_{r} \right) \Vec{r} \;,
\end{equation}
and energies as according to
\begin{equation}
  \tilde{E} = \left( \varepsilon_{r}^{2} / m^* \right) E \;,
\end{equation}
the Hamiltonian becomes
\begin{eqnarray}
  \tilde{\mathcal{H}}
  & = &
  - \frac{1}{2} \sum_{i=1}^{N} \tilde{\nabla}_{i}^{2}
  -  \sum_{i, I=1} ^{N}
  \frac{1}{\left| \tilde{\Vec{r}}_{i} - \tilde{\Vec{R}}_{I} \right|}
  \nonumber \\
  &&
  + \frac{1}{2}
  \sum_{i \neq j}^{N} \frac{1}{\left| \tilde{\Vec{r}}_{i} - \tilde{\Vec{r}}_{j} \right|}
  + \frac{1}{2}
  \sum_{I \neq J}^{N} \frac{1}{\left| \tilde{\Vec{R}}_{I} - \tilde{\Vec{R}}_{J} \right|}
  \label{eq:hamiltonian}
\end{eqnarray}
i.e. after rescaling the effective medium is replaced by the vacuum.
This means, for example, that the Mott criterion must be obeyed, though with a constant on
the right hand side that will be determined numerically below.
For easier comparison with experiment it is, however, convenient to retain the effective medium.
Below, we use the values
for silicon $m^* = 0.32$ and $\epsilon_{r} = 12$.

The next question is how to treat the model. To study a phase transition
we need to consider a reasonably large number of electrons, so
exact diagonalization is impractical.
Instead we use the Kohn-Sham formulation of
density functional theory \cite{hohenberg64, kohn65}.
The Kohn-Sham eigenfunctions $\psi_i$ and eigenvalues $\epsilon_i$
satisfy
\begin{equation}
  \left(
    - \dfrac{1}{2 m^*} \nabla^{2}
    + V_{\text{eff}}
    \right)
    \psi_i \left( \Vec{r} \right)
    = \epsilon_{i} \psi_i \left( \Vec{r} \right)
    \label{KS1}
\end{equation}
The effective potential
\begin{equation}
  V_{\text{eff}}
    =
    V_{\text{ext}}
    + V_{\text{Hartree}}
    + V_{\text{XC}}
\end{equation}
appearing in these equations is comprised
of three terms.
The first is the random potential due to the donors
\begin{equation}
 V_{\text{ext}} \left( \Vec{r} \right)
    =
    - \dfrac{1}{\varepsilon_{r}}
    {\displaystyle \sum_{I=1}^{N}}
    \dfrac{1}{\left| \Vec{r} - \Vec{R}_{I} \right|}
    \label{KS2}
\end{equation}
The second is the Hartree potential
\begin{equation}
    V_{\text{Hartree}} \left( \Vec{r} \right)
    =
    \dfrac{1}{\varepsilon_{r}}
    {\displaystyle \int} d^3 r^{\prime} \;
    \dfrac{n \left( \Vec{r}^{\prime} \right)}
    {\left| \Vec{r} - \Vec{r}^{\prime} \right|}
    \label{KS3}
\end{equation}
where
\begin{equation}
  n \left( \Vec{r} \right) =
  \sum_{i = 1}^{N} \left| \psi_i \left( \Vec{r} \right) \right|^{2}
  \label{KS4}
\end{equation}
is the number density of electrons.
The third is the exchange-correlation potential which is given by
the functional derivative of the exchange-correlation energy functional $E_{\text{XC}}[n]$
with respect to the number density
\begin{equation}
    V_{\text{XC}} \left( \Vec{r} \right) = \frac{ \delta E_{\text{XC}}}{\delta n \left( \Vec{r} \right) } \; .
    \label{KS5}
\end{equation}
If this functional were known exactly,
the solution of the Kohn-Sham equations would yield the exact ground state density and energy
of the interacting system.
In practice, the exact form of the
exchange-correlation energy functional is not known and an approximation is required.
In this work, we use the local density approximation (LDA)
\begin{equation}
    E_{\text{XC}} \approx E_{\text{XC}}^{\text{LDA}} \equiv
    \int d^3 r  \epsilon_{\text{XC}} \left( n \left( \Vec{r} \right) \right) n \left( \Vec{r} \right) \;.
\end{equation}
For $\epsilon_{\text{XC}}$ we use the form given in Eq. (2) of
Ref. \onlinecite{gunnarsson74},
with the parameter values of Ref. \onlinecite{janak75}.
Note that the expressions given in these references are for vacuum.
To use them for our effective medium we need to rescale lengths and energies as described previously.
The expression for the exchange-correlation energy becomes,
\begin{equation}
  E_{\text{XC}} \left[ n \right]
  =
  \frac{m_{e}^{*}}{\varepsilon_{r}^{2}} \cdot
  \tilde{E}_{\text{XC}} \left[ \tilde{n} \right]
\end{equation}
where the quantity with the tilde is the exchange-correlation energy found in the literature for vacuum.
To render the model more numerically tractable we assume complete spin polarization, i.e.
we set the spin polarization $\zeta = 1$

For numerical purpose we replace the continuous effective medium with
a cubic finite difference grid and
Eq. (\ref{KS1}) with a next-nearest neighbour finite difference approximation.
In the LDA the resulting matrix equations are sparse and the
eigenvectors and eigenvalues of the occupied states can be found using iterative techniques\cite{bollhofer07}.
The potential due to the impurities and the Hartree potential are found using fast Fourier transform.
The self-consistent solution of the equations is found iteratively starting from an initial guess for
the Kohn-Sham eigenfunctions.
For details the reader is referred to Ref. \onlinecite{harashima12}.

\section{Density of States}

\begin{figure}
  \begin{center}
    \includegraphics[width = \hsize]{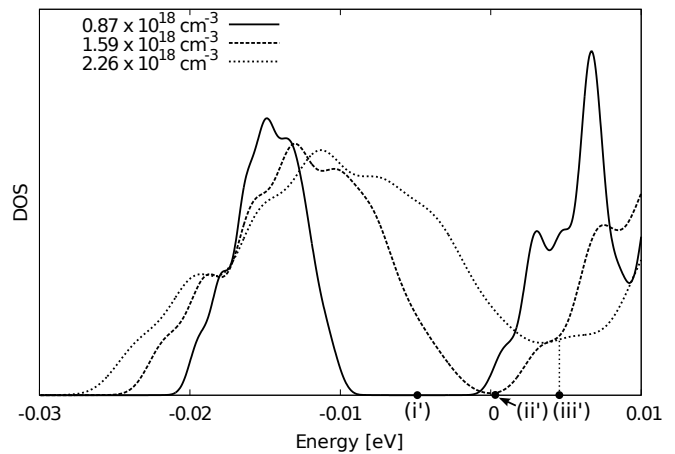}
  \end{center}
  \caption{The density of states for simple cubic systems.
    Points indicated by (i'), (ii'), and (iii') are Fermi levels
    for $N_{\text{d}} = 0.87$, $1.59$, and $2.26 \times 10^{18} \mbox{cm}^{-3}$,
    respectively.}
  \label{fig:dos-periodic}
\end{figure}

\begin{figure}
  \begin{center}
    \includegraphics[width = \hsize]{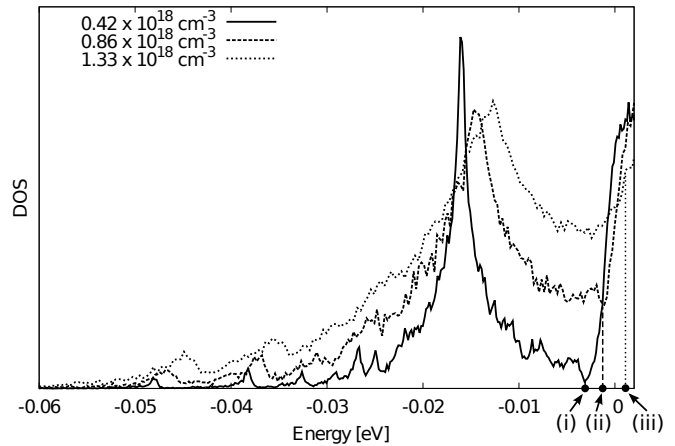}
  \end{center}
  \caption{The average density of states for a random spatial distribution of
    donor impurities.
    Points indicated by (i), (ii), and (iii) are Fermi levels
    for $N_{\text{d}} = 0.42$, $0.86$, and $1.33 \times 10^{18} \mbox{cm}^{-3}$,
    respectively.}
  \label{fig:dos-disordered}
\end{figure}

We take the density of Kohn-Sham eigenvalues as an approximation to
the single particle density of states.
For the unphysical situation that the donors are regularly
arranged on a simple cubic lattice, we have found that
the impurity band merges with the conduction band at $N_{\text{d}} \approx 1.59 \times 10^{18} \mbox{cm}^{-3}$ (Fig.~\ref{fig:dos-periodic}).
The average DOS for the more realistic random distribution of donors is shown, for several concentrations,
in Fig.~\ref{fig:dos-disordered}.
In this case, the bands merge at a much smaller concentration of
$N_{\text{d}} \approx 0.42 \times 10^{18} \mbox{cm}^{-3}$.
We should note this value is not directly comparable with experiment for two reasons.
First, we have assumed complete spin polarization.
Second, the LDA is known to underestimate band gaps.
Nevertheless, we can be sure that the MIT we observe below occurs after the impurity and conduction bands
have merged.

In the metallic regime, an anomaly is expected in the density of states at the Fermi level.
This anomaly is then expected to develop into a pseudo-gap (called the Coulomb gap) on passing far into the
insulating regime\cite{efros75}.
This behaviour has been seen clearly in a recent numerical analysis using the
Hartree-Fock method of a lattice model incorporating both disorder and Coulomb interactions\cite{amini14}.
For the model we consider here, the Coulomb gap will not be observable until after the impurity band starts to merge with the
conduction band.
For lower concentrations the impurity band is fully occupied (full spin polarization) and separated from the
conduction band by a gap.
For the lowest concentration shown in Fig.~\ref{fig:dos-disordered}, which corresponds to the
concentration at which the merging of the impurity and conduction band begins, a feature which we associate with the Coulomb gap
is visible at the Fermi level.
For the next highest concentration a dip but not a pseudo-gap can be seen.
At this concentration the highest occupied Kohn-Sham orbital is still localised.
For the highest concentration shown, which corresponds to a concentration where the highest occupied Kohn-Sham orbital is delocalised no
anomaly is discernable.


\section{multi-fractal finite size scaling analysis}
We observe a localisation-delocalisation transition of the highest occupied
Kohn-Sham eigenfunction as a function of donor concentration.
Before describing this, however,
We now turn to the multi-fractal finite size scaling analysis.
We divide the $L\times L \times L$ system
into boxes of linear size $l$.
We define a coarse grained intensity $\{\mu_{k}\}$ by
\begin{equation} \label{eq:mu}
  \mu_{k} \equiv
  \int_{k} d^{3}r \left| \psi \left( \Vec{r} \right) \right|^{2} \;.
\end{equation}
The subscript, $k$, indexes the $\left(L/l\right)^3$ boxes.
To analyze the transition we focus on the intensity distribution of the
highest occupied Kohn-Sham orbital.
We define
a generalized inverse participation ratio by
\begin{equation} \label{eq:r_q}
  R_{q} \equiv \sum_{k} \left( \mu_{k} \right)^{q}
\end{equation}
and the related quantity obtained by its differentiation with respect to the exponent $q$
\begin{equation} \label{eq:s_q}
  S_{q} \equiv \sum_{k} \left( \mu_{k} \right)^{q} \ln \mu_{k}
\end{equation}
Generalized multi-fractal exponents $\tilde{\tau}_{q}$ and $\tilde{\alpha}_{q}$
are defined from these
\begin{equation}
  \tilde{\tau}_{q}
  \equiv
  \frac{\ln \left< R_{q} \right>}{\ln \lambda}
  \; \; \; \; \; \; \; \; \; \;
  \tilde{\alpha}_{q}
  \equiv
  \frac{\left< S_{q} \right>}{\left< R_{q} \right> \ln \lambda}
\end{equation}
Here, $\lambda$ is the ratio of
box size to  system size
\begin{equation} \label{eq:lambda}
  \lambda \equiv \frac{ l }{ L }
\end{equation}
In the limit $\lambda\rightarrow 0$ the generalised exponents become the standard
multi-fractal exponents.

We proceed by assuming that, in the vicinity of the critical donor concentration,
the generalised multi-fractal exponents obey the scaling law\cite{rodriguez11}
\begin{equation} \label{eq:scaling-function-1}
  \Gamma \left( N_{\text{d}}, L, l \right)
  =
  F
  \left( \phi_1  L^{1/\nu}, \phi_2  L^{y_{2}}, \lambda \right)
\end{equation}
Here, $\Gamma$ indicates $\tilde{\tau}_{q}$ or $\tilde{\alpha}_{q}$,
$\phi_1$ is a relevant scaling variable, and $\phi_2$ an
irrelevant scaling variable with irrelevant exponent $y_{2}<0$.
(For brevity, the dependence on $q$ is not written explicitly.)
The scaling variables are functions of the reduced impurity donor concentration
\begin{equation} \label{eq:relative-concentration}
  n_{\text{r}} \equiv \frac{N_{\text{d}} - N_{\text{c}}}{N_{\text{c}}}
\end{equation}
To apply standard finite size scaling we fix
$\lambda$ at a constant value, i.e. we scale the box size simultaneously with the
system size so that we can omit the dependence on $\lambda$
\begin{equation} \label{eq:scaling-function-2}
  \Gamma \left( N_{\text{d}}, L \right)
  =
  F
  \left( \phi_1  L^{1/\nu}, \phi_2  L^{y_{2}} \right)
\end{equation}
The correlation length, $\xi$, is given by
\begin{equation} \label{eq:correlation-length-concentration}
  \xi = \xi_0 \left| \phi_1 \left( n_{\text{r}} \right) \right|^{- \nu}
\end{equation}
with $\xi_0$ a constant.
We allow for non-linearity of $\phi_1$ and $\phi_2$
by expanding them as
\begin{equation}
  \phi_1 \left( n_{\text{r}} \right) = \sum_{i = 1}^{m_1} a_{i} n_{\text{r}}^{i}
  \; \; \; \; \; \; \; \; \; \;
  \phi_2 \left( n_{\text{r}} \right) = \sum_{i = 0}^{m_2} b_{i} n_{\text{r}}^{i}
\end{equation}
At the critical point $\phi_1$ must be zero, so we fix the constant term in the
expansion to zero.
The scaling function Eq.~(\ref{eq:scaling-function-2}) is expanded as
\begin{equation} \label{eq:scaling-function-expanded}
  F \left( X, Y \right)
  = \sum_{i, j = 0}^{n_1, n_2} F_{i j} X^{i} Y^{j} \;.
\end{equation}
We estimate the critical concentration and the critical exponent by fitting this model to
our simulation data.

This method has previously been
successfully applied to the Anderson transition\cite{rodriguez10,
rodriguez11}.
In addition, two recent papers\cite{burmistrov13,amini14} have shown that multi-fractality survives in the presence of the Coulomb interaction.

\section{Metal-Insulator Transition}

Simulations were performed for system sizes in the range $L = 229 \sim 400 \mbox{\AA}$
and donor concentrations of $N_{\text{d}} \equiv N / L^{3} = 0.4 \sim 1.3 \times 10^{18} \mbox{cm}^{-3}$.
This corresponds to 5 $\sim$ 85 electrons.
We set the finite difference grid spacing to 18 Bohr, which is about half of the effective Bohr radius for Si.
The donors were randomly distributed on a simple cubic lattice with
spacing 36 Bohr.
This avoids the
situation that two donors are unphysically close.
The number of samples for each system size and donor concentration
varies between $1500$ and $3000$.
For a few percent of samples,
the self-consistent calculation does not converge.
These samples are neglected in the analysis.
For the multi-fractal analysis we set $\lambda = 1/6$.

Fig.~\ref{fig:alpha0} shows the generalised exponent
$\tilde{\alpha}_{0}$ as a function of the donor concentration.
For low concentration
$\tilde{\alpha}_{0}$ increases with system size.
This is the typical behavior for localized states.
Opposite behavior, typical of extended states, is found for higher concentration.
The scale invariant point between these concentration regions indicates the transition.
The solid lines in Fig.~\ref{fig:alpha0} are the result of fitting
the finite size scaling model to the data.
The number of data points is 67.
The orders of the expansions are
$m_1 = 2$, $m_2 = 1$, $n_1 = 3$, $n_2 = 1$ and
the number of corresponding fitting parameters is 13.
The critical concentration and critical exponent obtained from the fitting
are $N_{\text{c}} = 1.09 (+0.07, -0.01) \times 10^{18}\mbox{cm}^{-3}$ and
$\nu = 1.30 (+0.12, -0.06)$.
The errors are 95\% confidence intervals obtained by Monte Carlo simulation.

\begin{figure}[h]
  \begin{center}
    \includegraphics[width = \hsize]{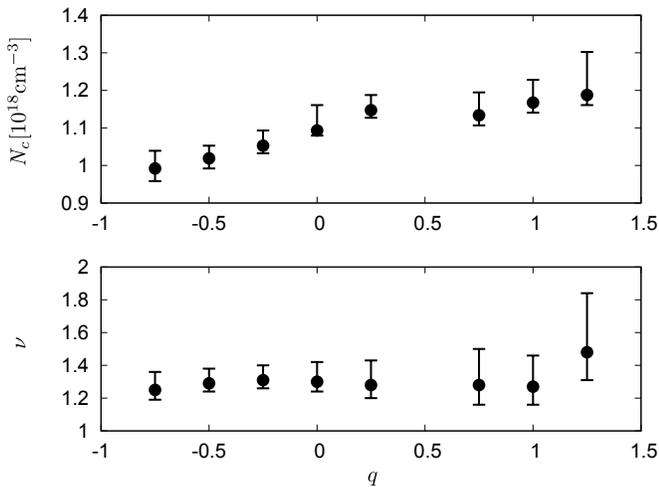}
  \end{center}
  \caption{The estimates of the critical concentration $N_{\text{c}}$ and the
  critical exponent $\nu$ obtained for different powers $q$.
  The data shown above is estimated from $\alpha_{q}$ and
  similar behaviour can be observed for $\tau_{q}$.}
  \label{fig:q}
\end{figure}

The estimates of the critical concentration and the critical exponent should not depend on $q$.
To check this,
we performed the multi-fractal analysis for various for $q = -0.75 \sim 1.25$.
The results are shown in Fig.~\ref{fig:q}.
On the one hand, the estimate of the critical concentration varies by about 20\%
for the range of $q$ considered, which is not completely satisfactory.
On the other hand, the estimate of the critical exponent is  independent of $q$.

The measured critical concentration for Si:P is $3.52 \times 10^{18} \mbox{cm}^{-3}$.
The predicted critical concentration based on the Mott criterion is
$2.25 \times 10^{18} \mbox{cm}^{-3}$ and we should certainly not expect to do better
than this in a theoretical approach that uses an effective medium.
The MIT in our model occurs at $1 \sim 1.2 \times 10^{18}\mbox{cm}^{-3}$
which corresponds to a value of 0.2 in the Mott criterion.
Possible explanations for this discrepancy
include the neglect of the spin degree of freedom and limitations of the LDA.

Our result for the critical exponent, $\nu \approx 1.3$, should have
a wide applicability since the values of critical exponents are universal.
While our range of system sizes is limited, the $q$ independence of the estimate of
the exponent seen in Fig.~\ref{fig:q} gives us some confidence that the critical exponent is
different from that of the standard Anderson model.
This is puzzling because the Kohn-Sham equations appear to be in the same
Dyson symmetry class as Anderson's model of localisation.
In the local density approximation there is no long range hopping. And while
the long range nature of the Coulomb interaction leads to spatial correlations in
the diagonal matrix elements of the Kohn-Sham Hamiltonian,
a numerical investigation showed that these correlations appear to be exponential with a decay length roughly equal to
the Thomas-Fermi screening length.

\section{Discussion}

We estimated the critical concentration and critical exponent of
the metal-insulator transition in a simulation of a doped semiconductor.
Our simulation incorporates both the disorder due to the random positions of the
donor impurities and the Coulomb interaction between electrons.
We took account of the Coulomb interactions 
using density functional theory in the local density approximation, and used
multi-fractal finite size scaling to analyse the metal-insulator transition.
We found that the correlation length critical exponent appears to be
different from
that of the non-interacting Anderson model of localisation.
This suggests that the Coulomb interaction changes
the universality class.
This is consistent with the analytic study reported in Ref. \onlinecite{burmistrov13} where it was found that
the multi-fractal spectrum differs from that in the absence of interaction.
Our estimate of the critical exponent differs from the obtained in a study of a lattice model within the Hartree-Fock approximation reported
recently in Ref. \onlinecite{amini14}.
Those authors found different values of the correlation exponent on the insulating and metallic sides of the transition,
something which is very difficult to reconcile with our results.

In principle, density functional theory is exact and gives the energy and density of the many
body ground state.
In practice, of course, density functional calculations are approximate.
The principle approximations in this work are two-fold.
The first is the use of the local density approximation.
It might be possible to improve on this by the use of generalized gradient approximations.
The second is the use of the Kohn-Sham wavefunctions in place of the true many-body
ground state wavefunction.
This might be addressed by using the
current calculation as a starting point for many-body perturbation calculations.
The current study could also be extended in several other simpler ways.
It should be possible to study the role of compensation by introducing
negatively charged donors.
The restriction of complete spin polarization may be relaxed
by using the local spin density approximation, which would allow
issues related to local moments to be addressed
\cite{bhatt92, belitz94, finkelstein10, kettemann09, kettemann12, potter12}.

\begin{acknowledgments}
We thank Stefan Kettemann and Tomi Ohtsuki
for critical readings of this manuscript.
KS thanks  the Advanced Materials Science division of POSTECH and Jacobs University, Bremen for their hospitably during visits in the course of this work.
Part of the computations were performed using
the Supercomputer Center, ISSP,
the University of Tokyo.
YH acknowledges the support of the Global COE Program (Core Research and
Engineering of Advanced Materials-Interdisciplinary Education Center for Materials
Science), MEXT, Japan.
\end{acknowledgments}

\bibliography{Reference}

\end{document}